\documentclass[12pt,onecolumn,draftcls]{IEEEtran}

\usepackage{cite,graphicx,array,epsf,epsfig,color,subfigure,ulem}
\begin{document}

\title{Theory of high bias Coulomb Blockade in ultrashort molecules}
\author{Bhaskaran~Muralidharan$^1$,
  Avik~W.~Ghosh$^2$,~\IEEEmembership{Member,~IEEE,}, 
~Swapan~K.~Pati$^3$, and
~Supriyo~Datta$^{1}$,~\IEEEmembership{Fellow,~IEEE}
\\[10pt]\small
\begin{tabular}[t]{c} $^1$ School of Electrical and Computer Engineering and Network for Computational Nanotechnology
    \\ Purdue University \\ West Lafayette, IN 47907, USA \\ Email: {\{bmuralid\}@ecn.purdue.edu}\\
 $^2$ School of Electrical and Computer Engineering
    \\ University of Virginnia \\ Charlottesville, VA 22904, USA  \\ 
 $^3$ Theoretical Sciences Unit and Chemistry and Physics of Materials Unit
    \\ Jawaharlal Nehru Center for Advanced Scientific Research, Jakkur Campus  \\Bangalore 560064, India  \\ 
 \end{tabular}
}

\maketitle

\begin{abstract}
We point out that single electron charging effects such as Coulomb Blockade (CB) and 
high-bias staircases play a crucial role in transport through single ultrashort 
molecules. A treatment of Coulomb Blockade through a prototypical molecule, benzene, is developed using a 
master-equation in its complete many-electron Fock space, evaluated through exact diagonalization or 
full Configuration Interaction (CI). This approach can explain a whole class of non-trivial experimental features including vanishing zero 
bias conductances, sharp current onsets followed by ohmic current rises, and gateable current 
levels and conductance structures, most of which cannot be captured 
even qualitatively within the traditional Self Consistent Field (SCF) approach coupled
with perturbative transport theories. By comparing the two approaches, namely SCF and CB, 
in the limit of weak coupling to the electrode, we establish that the 
inclusion of strong-correlations within the molecule becomes critical in addressing the above experiments. 
Our approach includes on-bridge-correlations fully, and is therefore well-suited
for describing transport through short molecules in the limit of weak coupling to electrodes.
\end{abstract}


% Will be ignored for other modes.
\IEEEpeerreviewmaketitle

\section{Introduction}
  Theoretical calculations on single molecule conduction have typically employed coherent
  Non-Equilibrium Green's function (NEGF) theories (``Landauer limit") \cite{rdatta1,mwl1} coupled with Self Consistent Fields (SCF) to describe charging effects.
  Though fairly successful in describing many aspects of single molecule conduction \cite{diven,damle,jt,rmrs,rasymm}, there have been important discrepancies
between theory and experiment \cite{rreed}. The most common ones include
poor match between theoretical and experimental current levels and zero-bias currents \cite{diven,damle,rreed}.
It was also pointed out in \cite{rbhasko} that a whole class of experimental I-V's show features,
which cannot be captured even qualitatively using an SCF theory.
Charging energies of short molecules ($~3$ eV for benzene) are often larger than their electrode coupling 
($< 0.2 $ eV for benzene di-thiol on gold),
and thus could be in the Coulomb Blockade (CB) regime where single electron charging effects could dominate.
It is thus debatable whether it is better described as a quantum wire in the SCF regime, or as a quantum dot array 
in the Coulomb Blockade (CB) regime. Nevertheless the wisdom of SCF approaches must be 
scrutinized especially for conduction through shorter molecules. The purpose of this paper 
is to present a Coulomb Blockade approach to
molecular conduction using a benzene molecule as prototype, and establish it 
as a different viewpoint from the conventional NEGF-SCF treatment. Furthermore features obtained via 
the CB approach can semi-quantitatively explain several
non-trivial features commonly observed \cite{jpark,rweber1,rweber2,pnas,rscott} in experiments.

\begin{figure}[ht]
\hskip 0.7cm\centerline{\epsfxsize=6.0in\epsfbox{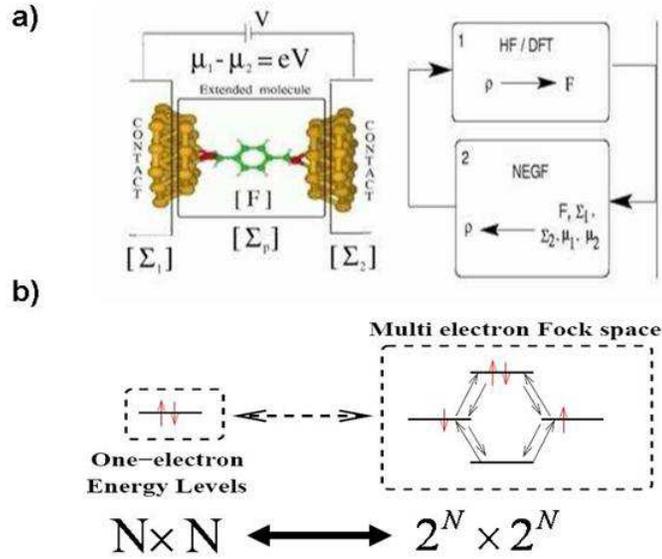}}
\caption{Our system is a benzene molecule coupled to metallic contacts.
Single molecule transport calculations typically employ the NEGF-SCF prescription. 
The Block diagram depicts the basic scheme. While quantities such as the Hamiltonian are
in the one-electron space of dimensions $N \times N$, N being the number of
basis functions. Our Coulomb Blockade description 
involves the full many electron Fock space of dimensions $2^N \times 2^N$ as shown in b) 
using a single spin-degenerate level as an example. The use of full many-electron space 
captures the correlations exactly within the framework of the given 
$N \times N$ one-electron Hamiltonian.}
\label{fig_sim}
\end{figure}

   It is common to distinguish between two regimes of transport: a) an SCF regime where 
the dominant
energy scale is the contact coupling, allowing for fractional charge transfer through the system;
and b) a Coulomb Blockade (CB) regime where the dominant energy scale is the single electron charging, 
leading to integral charge transfer. In the SCF regime the description of transport via
non-interacting single particle energy levels can be justified. In this limit, it is common to
use the SCF-NEGF scheme that takes charging effects into account. 
Here the molecular Hamiltonian is described by a set of single particle levels,
which are coupled to reservoirs through their self energies. The electron interactions are taken into account using SCF schemes as shown in the block diagram in Fig. 1a.
All quantities in the NEGF formalism are matrices of dimension $N \times N$, $N$
being the number of single particle basis functions used. This allows for an accurate
description of quantum chemistry of both the isolated molecule and its bonding to the contacts \cite{liang}. 
In the CB regime with weak contact coupling, charging effects dominate, and the use of single 
particle basis sets may be questionable. In such cases, it may be preferential to employ a multi-electron description or 
Configuration Interaction (CI) where feasible. The central quantities in this CI method are now matrices of 
dimension $2^N \times 2^N$, thereby accounting for strong interaction accurately. The weakly coupled contacts are treated perturbatively 
using transition rates between states differing by a single electron [5].
It is interesting to note that most theoretical efforts in molecular conduction have
been in the SCF regime, while energy scales favor the CB regime. Our paper is thus a concrete attempt 
towards CI based transport.

This paper is organized as follows: we begin by defining an appropriate many-body Hamiltonian for 
Benzene whose parameters are benchmarked based on well-established mean-field techniques.
We then illustrate how a CB treatment is conceptually different from the standard SCF treatment in the weak coupling limit, not only
under non-equilibrium conditions, but even under equilibrium conditions. We then point out the importance of 
inclusion of excited states in transport, that naturally arise 
within our CI approach. The progressive access of these excited states leads to transport signatures under various 
non-equilibrium conditions. Before we conclude, a few CB fits to experimental data are presented in 
support of our analysis.

\section{The Model Hamiltonian and Equilibrium Properties} 
An appropriate model Hamiltonian is usually described with an adequate basis set.
In this paper, we use a tight binding Hamiltonian with one $p_z$ orbital per site to describe our CI based scheme.
Although this generates just a minimal $6 \times 6$ single particle basis set, its many-electron space is $2^{12} \times  2^{12}$ in size.
Besides, our objective here is to describe the CI approach for transport and compare it with
the SCF approach for the same Hamiltonian. Better quantum chemical descriptions within the CI approach can be achieved by starting 
with a reduced but more accurate one-particle Hamiltonian, but we leave these for future work.

One begins with the model Hamiltonian in second quantized notation:
\begin{eqnarray}
\hat{H} &=& \sum_{\alpha} \epsilon_{\alpha} n_{\alpha}
+ \sum_{\alpha \neq \beta} t_{\alpha \beta} c_{\alpha}^{\dagger} {c_\beta} \nonumber\\
&+& \sum_{\alpha,\sigma} U_{\alpha \alpha} n_{\alpha\sigma}
n_{\alpha\bar{\sigma}} + \frac{1}{2} \sum_{\alpha \neq \beta} U_{\alpha \beta}
n_{\alpha} n_{\beta} ,
\label{eq:mbh}
\end{eqnarray}
where $\alpha,\beta$
correspond to the orbital indices of the frozen $p_z$ orbitals for carbon sites on the Benzene 
ring,and $\sigma$,$\bar{\sigma}$ represent a particular spin and its reverse.
In connection to its equilibrium configuration, it is more convenient to work with onsite
energies $\tilde{\epsilon}$ defined as:
\begin{equation}
\tilde{\epsilon}_{\alpha}
= \epsilon_{\alpha} + U_{\alpha \alpha} \langle n_{\alpha \bar{\sigma}}\rangle + \frac{1}{2}
\sum_{\alpha \neq \beta} U_{\alpha \beta} \langle n_{\beta} \rangle ,
\label{eq:mf}
\end{equation}
where $\tilde{\epsilon}_{\alpha}$'s denote the
mean-field on-site energies in the equilibrium charge neutral configuration of the molecule 
and $\langle n\rangle$ represents its mean-field value. 
Now the model Hamiltonian is simply re-written as:
\begin{eqnarray}
\hat{H} &=& \sum_{\alpha} \tilde{\epsilon}_{\alpha}
n_{\alpha} + \sum_{\alpha \neq \beta} t_{\alpha \beta} c_{\alpha}^{\dagger}
{c_{\beta}} \nonumber\\&+& \sum_{\alpha,\sigma} U_{\alpha \alpha}
(n_{\alpha\sigma} - \langle n_{\alpha \sigma}\rangle) (n_{\alpha \bar{\sigma}} - \langle n_{\alpha
\bar{\sigma}}\rangle) \nonumber\\&+& \frac{1}{2} \sum_{\alpha \neq \beta} U_{\alpha
\beta} (n_{\alpha}-\langle n_{\alpha}\rangle) ( n_{\beta} -
\langle n_{\beta}\rangle).
\label{eq:mfh}
\end{eqnarray}
The mean-field Hamiltonian derived from the above Hamiltonian, is:
\begin{equation}
\hat{h}=\sum_{\alpha} \tilde{\epsilon}_{\alpha}
n_{\alpha} + \sum_{\alpha \neq \beta} t_{\alpha \beta} c_{\alpha}^{\dagger}
{c_{\beta}} + U^{SCF}_{\alpha \alpha},
\label{rscf1}
\end{equation}
where
\begin{equation}
U^{SCF}_{\alpha \alpha}= U_{\alpha \alpha}(n_{\alpha} -
\frac{\langle n_{\alpha}\rangle}{2}) + \frac{1}{2} \sum_{\alpha \neq \beta} U_{\alpha \beta}
(n_{\beta} - \langle n_{\beta}\rangle).
\label{eq:rscf}
\end{equation}
is the Self Consistent Field, the calculation of $\langle n_\alpha\rangle$  performed self consistently with 
the one electron Hamiltonian $\hat{h}$. In the following sections, we derive appropriate parameters 
$\tilde{\epsilon}$, $t$ and $U$ for benzene, to describe the two different approaches i.e., the CI (Eq.~\ref{eq:mfh}) and the 
SCF approaches (Eq.~\ref{eq:rscf}), and compare them in parallel in the case of both equilibrium and 
non-equilibrium conditions.

\begin{figure}
%\centerline{\epsfig{figure=fig2_n.ps,width=3.4in}}
\centerline{\epsfig{figure=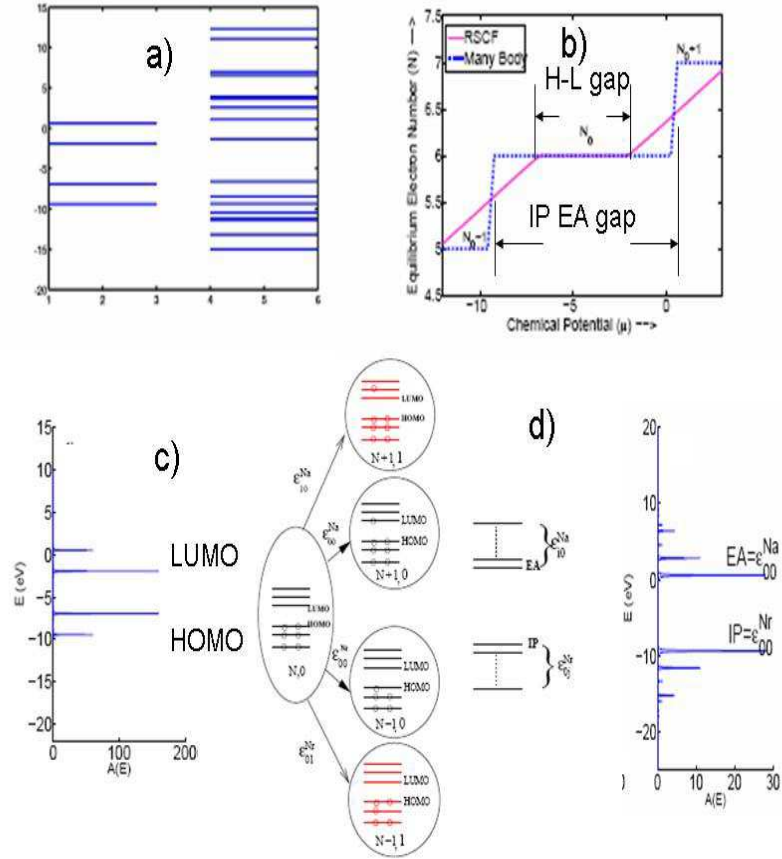,width=5in,height=5in}}
\caption{Model Hamiltonian and Equilibrium Properties. (a) Selection of on-site $\epsilon_{\alpha}$ 
and hopping parameter $t_{\alpha \beta}$. Comparison of our model Hamiltonian levels 
with frontier LDA/6-31g levels. Parameters are fixed based on a close match between 
the doubly degenerate HOMO and LUMO levels and singly degenerate HOMO-1, LUMO+1 levels. (b)
Charging parameter matched according to a consistent Restricted SCF based $N-\mu$ plot (shown continuous line). 
Total energy based Many-Body calculation (shown dotted line) as well as RSCF calculation is consistent 
with Gaussian based calculation \cite{rrak}. c) One particle spectral function shows 
peaks at the energy levels of the single particle Hamiltonian $\tilde{h}$. 
d) Lehmann spectral function evaluated via many-electron spectrum yields many 
more spectral peaks corresponging to removal (addition) of electrons 
from the neutral ground state into various charge configurations (excitations) of singly charged species. 
Notice that the IP-EA and HOMO-LUMO gaps are equal to the corresponding charge-stability 
plateaus $N=N_0$ for many-body and SCF calculations shown in b).}
\label{fig:ben}
\end{figure}

Fig.2(a) shows the selection of mean field on-site energies $\tilde{\epsilon}_{\alpha}$ and hopping
parameter $t_{\alpha \beta}$, by comparing the eigen-energies of our model SCF Hamiltonian (Eq.~\ref{rscf1})
with the frontier orbitals within the local density
approximation (LDA) in the 6-31g basis set, shown in the left and right sections of Fig. 2a respectively. 
The carbon-carbon hopping term $t_{\alpha \beta}=-2.0 eV$
has been used from already tabulated data \cite{ralbert}, which yields $\tilde{\epsilon}_{\alpha}=-4.42 eV$ for the above fit. 
Note that the Highest Occupied Molecular Orbital (HOMO) levels and Lowest Unoccupied Molecular Orbital (LUMO) 
levels are doubly degenerate in our model tight binding Hamiltonian as well as in the LDA basis set.

\subsection{Equilibrium Electron Number v/s chemical potential: Choosing Charging parameters}
A distinguishing aspect of CB is the abrupt charge addition as opposed to a gradual one in 
an SCF calculation shown in Fig.2(b). This fact is readily seen in the figure, 
in which the SCF and CB calculations are presented using the one electron and many-electron 
Hamiltonians described by Eq.~\ref{eq:mfh} and~\ref{rscf1} respectively. In the weak coupling limit, 
the CB result is more physical, and the SCF calculation does not do justice to this integer charge transfer. 
However, schemes such as self interaction correction could be introduced within the one-electron Hamiltonian 
\cite{rpal,ssan,pals} to incorporate this. But it turns out that even such schemes may not capture non-equilibrium correctly \cite{rbhasko}. It 
is however expected that as coupling strength to the electrodes is increased, 
the electron transfer resembles the SCF result. There is as yet no clear formalism that addresses \cite{rmat} 
this crossover, even in the equilibrium case although the two 
opposite limits namely SCF and CB are well understood. While the two limits can individually be handled
by perturbative expansion in the small parameters $U/\Gamma$ and $\Gamma/U$, $U$ being the single electron
charging energy and $\Gamma$ being the level broadening, the intermediate regime is hard to handle owing to
the non-existence of a suitable small parameter or `fine structure constant' for transport.

The plateau in the charge addition diagram $N$ versus $\mu$, in which the electron number is stabilized, 
spans the HOMO-LUMO gap in the SCF case and the Ionization potential-Electron Affinity (IP-EA gap) in the 
CB calculation. The IP (EA) is defined as the energy when an electron can be removed (added) to the neutral 
molecule carrying $N_0$ electrons. This occurs when the chemical potential $\mu$ equals the energy difference between 
ground states differing by an electron number $\mu=E_G^{N_0}-E_G^{N_0-1}$ for IP, 
and $\mu=E_G^{N_0+1}-E_G^{N_0}$ for EA. The situation is however different in the case of SCF. 
Here the charge transfer dictated by a self consistent potential (Eq.~\ref{eq:rscf}) is gradual, 
in which two electrons are transferred adiabatically over a span of $2U$ corresponding to the removal of two electrons. 
This is usually referred to as the restricted SCF (RSCF mentioned in Fig. 2b). Most SCF calculations 
in the literature \cite{rfulde} employ different variants of this scheme. There are also 
spin unrestricted SCF techniques \cite{ssan,rpal,pals} which take into account the 
abrupt charge transfer in a weakly coupled system, due to self-interaction correction, 
but it is not yet clear whether they work out of equilibrium \cite{rbhasko,rbhasko2}. 

One expects that the IP occurs roughly midway during the gradual charge removal in the RSCF scheme \cite{rmat,rfulde}.  
We use this fact to estimate our charging parameters $U_{\alpha \beta}$, with the aid of a Gaussian-98 based
calculation for the equilibrium electron number v/s chemical potential ($N-\mu$),
published elsewhere \cite{rrak}. The calculation corresponding to the equilibrium number of 
$N_0=42$ maps onto our model calculations for $N_0=6$, focussing thus on the frontier
orbitals and ignoring the inner core that is frozen in our estimate 
for $\tilde{\epsilon}_\alpha$. 
By implementing a Restricted SCF scheme using Eq. ~\ref{eq:rscf} within in our
model Hamiltonian, we obtain a close match of the $N-\mu$ plots in the range between $N_0$
and $N_0-1$ in comparison with the Gaussian-98 calculation in \cite{rrak}. Using an estimate of the onsite 
charging $U_{\alpha \alpha}$, we calculate $U_{\alpha \beta}$ using the Matago-Nishimoto
approximation:
\begin{equation}
U_{\alpha \beta}=\frac{e^2}{4\pi\epsilon_0
r_{\alpha \beta} +\frac{2e^2}{U_{\alpha \alpha}+U_{\beta \beta}}},
\label{eq:ppp}
\end{equation}
where $r_{\alpha \beta}$ is the inter carbon distance in benzene. In 
each case, evaluation of $n_{\alpha}$ is done self-consistently using an equilibrium value $N_0=6$, and 
$\langle n_{\alpha \sigma}\rangle=\frac{1}{2}$. 
Using exact eigen-energies of the many-electron Hamiltonian Eq.~\ref{eq:mfh} with the above parameters, 
an $N-\mu$ calculation using these total energies (shown dotted red in Fig.2b) 
is in excellent agreement with respect to Gaussian calculations in \cite{rrak}. 
Note that the $\mu=IP$ in Fig.2b occurs midway between $N=N_0$ and $N=N_0-2$ in the RSCF charging diagram. 
It is worth mentioning that the many-body calculation presented 
in this figure takes all correlation energies into account and is the exact ground state 
energy within our defined model Hamiltonian. 
\subsection{Equilibrium Spectral function}
Conduction through molecules via molecular orbitals is well understood in the SCF picture \cite{rfer}. 
In the strongly coupled regime (most appropriate for an SCF treatment), fractional
charge transfer occurs, and Density of States (DOS) is evaluated at
equilibrium \cite{rdatta1} in order to capture the effect of the strong coupling
with contact. An interplay of molecular DOS and charging treated self-consistently  
determines the non-equilibrium response (current-voltage or I-V characteristics). 
The density of states calculated from the one-electron Green's function \cite{rdatta1} 
in Fig. 2c shows peaks at the single electron eigen spectrum. 
As the coupling to electrodes gets stronger, the single electron DOS will show signatures and 
artifacts of contact bondings \cite{liang,rfer}. 

In the weak coupling (CB) limit however, integer charge addition is favored, and 
transitions between states that differ by a single electron appear as spectral signatures \cite{rralph}. At
equilibrium, it is convenient to introduce the {\it{Ground State Spectral
Function}} by defining the Green's function in the Lehmann
representation \cite{rfulde}:
\begin{eqnarray}
G_{\alpha \beta}(E) &=& \frac{\langle N,0|c_{\alpha}|N+1,j\rangle\langle N+1,j|c^\dagger_{\beta}|N,0\rangle}{E+i0^+ - (E^{N+1}_j -E^N_0)} \nonumber\\
&+& \frac{\langle N,0|c^{\dagger}_{\beta}|N-1,j\rangle \langle N-1,j|c^{}_{\alpha}|N,0\rangle}{E+i0^+ - (E^N_0 - E^{N-1}_j)} \nonumber\\
A_{\alpha \beta}(E)&=& i[G(E)-G^\dagger (E)] 
\label{eq:leh}
\end{eqnarray}
where $\alpha, \beta$ correspond to the orbital index, 
which in our case are the sites of the benzene molecule, and $| N, i \rangle$ 
denotes the $i^{th}$ excited state of 
a charge configuration of $N$ electrons. The poles of this spectral function represent various
transition energies for addition (removal) of electrons from the neutral ground
state:
\begin{eqnarray}
\epsilon^{Nr}_{0j}&=& E^{N}_{0}-E^{N-1}_{j} \nonumber\\
\epsilon^{Na}_{0j}&=& E^{N+1}_{j}-E^{N}_{0}
\label{eq:sp0}
\end{eqnarray}
whose spectral strengths are given by:
\begin{eqnarray}
{\tau}^{Nr}_{0j,\alpha \beta} &=& \langle
N,0|c_{\alpha}|N+1,j\rangle\langle N+1,j|c^\dagger_{\beta}|N,0\rangle \nonumber \\
{\tau}^{Na}_{0j,\alpha \beta} &=& \langle
N,0|c^{\dagger}_{\beta}|N-1,j\rangle\langle N-1,j|c^{}_{\alpha}|N,0\rangle
\label{eq:w0}
\end{eqnarray}
The first (addition) term adds an electron to orbital $\beta$, taking the system from an N electron
ground state to the $j^{th}$ (N+1) electron excited state, and then removes it from orbital $\alpha$, bringing
it back to ground state. The second (removal) equation first removes an electron from $\alpha$ and then adds it to
$\beta$. 

One can re-write the expression in terms of {\it{diagonal terms only}}, replacing the recurring index $\alpha$ with a single index, in a more convenient form as:
\begin{equation}
A^N_{0\alpha}(E)=\sum_{j}\left[{\tau}^{Nr}_{0
j \alpha}\delta(E-\epsilon^{Nr}_{0j})+{\tau}^{Na}_{0j
\alpha}\delta(E-\epsilon^{Na}_{0j}) \right]
\label{eq:sp0}
\end{equation}
The spectral function shown in Fig. 2d represents the removal and addition strength of various transitions at 
their energies given by Eq.~\ref{eq:sp0}. Notice that there are numerous peaks in this spectrum calculated 
from the many-electron transitions, due to the possible transfer to 
various excited states of charged species shown in Fig. 2d. It is 
important to note that although each transition has a non-trivial spectral weight given by Eq.~\ref{eq:w0}, 
{\it{they satisfy an overall sum rule}} that amounts to the total electron number in the system. We will 
see in 
subsequent sections 
that these transitions involving excited states show up directly as transport signatures frequently 
observed in experiments.
\section{Non-Equilibrium}
This section is devoted to the various unique transport signatures in the weak coupling (CB) regime, many 
of 
which have experimental significance. We elaborate on how various excited states get accessed as a result 
of contacts maintained at different potentials (non-equilibrium), and what the SCF theory completely 
misses in this regime. Throughout this paper we describe the electrodes (contacts) 
using corresponding electrochemical potentials $\mu_L$ and $\mu_R$ and coupling strengths $\gamma_L$ and $\gamma_R$.

\subsection{Coulomb Blockade approach: Rate equation model}
Transport in the CB limit \cite{rlikharev,rralph,rhettler} is often modeled with a rate equation approach, 
in which the steady state addition and removal of electrons is described with a rate equation for the
nonequilibrium probability $P^N_i$ of each N electron many-body state
$|N,i\rangle$ with total energy $E^N_i$. The master equation 
involves transition rates $R_{(N,i)\rightarrow(N\pm 1,j)}$ between states differing by a 
single electron, leading to a set of independent equations defined by the size of the
Fock space \cite{rralph}
\begin{equation}
\frac{dP^N_i}{dt} =
-\sum_j\left[R_{(N,i)\rightarrow(N \pm 1,j)}P^N_i -R_{(N\pm
1,j)\rightarrow(N,i)}P^{N\pm 1}_j\right]
\label{ebeenakker}
\end{equation}
along with the normalization equation $\sum_{i,N}P^N_i = 1$. We define rate constants
\begin{eqnarray}
\Gamma_{ij\alpha}^{Nr} &=& \gamma_\alpha|{\tau}^{Nr}_{ij\alpha}|^2\nonumber\\
\Gamma_{ij\alpha}^{Na} &=& \gamma_\alpha|{\tau}^{Na}_{ij\alpha}|^2,
\end{eqnarray}
where $\gamma_{\alpha}$
represents lead molecule broadening or coupling via the end atoms, described
using Fermi's Golden rule. These constants represent
the partial probability for the electron to be injected by the end atom 
into a given many-electron ground or excited state. The transition rates are
now given by
\begin{eqnarray}
R_{(N,i)\rightarrow(N-1,j)} &=& \sum_{\alpha=L,R}\Gamma_{ij\alpha}^{Nr}\left[1-f(\epsilon^{Nr}_{ij}-\mu_\alpha)\right]\nonumber\\
R_{(N-1,j)\rightarrow(N,i)} &=&\sum_{\alpha=L,R}\Gamma_{ij\alpha}^{Nr}f(\epsilon^{Nr}_{ij}-\mu_\alpha).
\end{eqnarray}
for the removal levels $(N,i \leftrightarrow N-1,j)$, and replacing $(r
\rightarrow a,   f \rightarrow1-f)$ for the addition levels $(N,i \leftrightarrow N+1,j)$. $\mu_\alpha$ are the contact electrochemical potentials, $f$ is the
corresponding Fermi function, with single particle removal and addition energies
$\epsilon^{Nr}_{ij} = E^N_i - E^{N -1}_j$, and $\epsilon^{Na}_{ij} = E^{N+1}_j -
E^N_i$. Finally, the steady-state solution to Eq.(\ref{ebeenakker}) is used to
get the left terminal current as
\begin{equation}I =
\pm\frac{e}{\hbar}\sum_{ij}\left[R^L_{(N,i)\rightarrow(N\pm 1,j)}P^N_i-
R^L_{(N\pm 1, j)\rightarrow(N,i)}P^{N\pm 1}_j \right]
\end{equation}
where states
corresponding to a removal of electrons by the left electrode involve a negative
sign. 
\subsection{STM Limit:}
We briefly elucidate the relationship between spectral functions defined earlier and STM conduction spectra.
 Electrical conduction depends on the measurement geometry \cite{rfer} and charging determined by 
{\it{capacitive}} voltage-division ratio $\eta$ between leads in the Laplace solution as
opposed to the
{\it{resistive}}
voltage-division ratio $\gamma = \gamma_R/\gamma_L$ which determines the extent
to which the levels are filled or emptied by the leads. The source and
drain potentials are then given by $\mu_L = E_F + \eta V_d$ and $\mu_R = E_F -(1-\eta)V_d$.

Consider a simple picture in which one contact is very weakly coupled ($\gamma<<1, \eta=0$), 
equivalent to the molecule being in equilibrium with left contact $\mu_L$. In the $\eta=0$ 
limit the molecular energy levels are pinned to this contact implying that for a positive voltage $\mu_R < \mu_L$, 
and $\mu_L$ remains at the equilibrium position. This picture is analogous to STM {\it{shell tunneling}} experiments \cite{rbanin}, in which the weakly 
coupled STM tip acts as a voltage probe, thereby generating the
single particle spectrum, the molecule/dot held in equilibrium with the more
strongly coupled contact, in this case, the substrate.
\begin{itemize}
\item {\it{Ground State Spectral function}}: It is expected, with a more strongly coupled contact, 
that the right contact voltage probe, such as an STM tip can add or withdraw into or out of the 
dot at energies corresponding to addition or removal energies defined in Eq. ~\ref{eq:spl}. The 
stronger coupling to the left contact ensures that an electron be added or removed as soon as the tip removes
or adds an electron thus maintaining overall charge neutrality. In this case
conductance spectrum proportional to the equilibrium spectral function is
obtained as shown in Fig.3b.
\begin{figure}
\centerline{\epsfig{figure=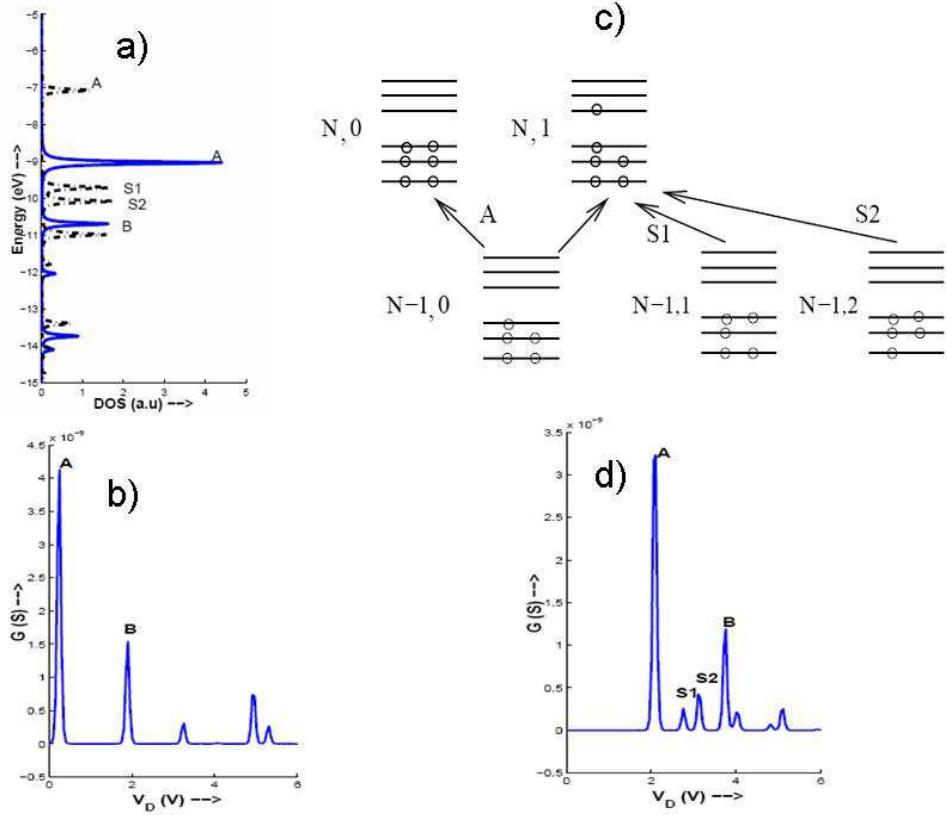,height=4.5in,width=6in}}
%\centerline{\epsfxsize=4in\epsfbox{CB_fig3.ps}}
\caption{STM limit - mapping spectral signatures: a) Removal Spectral functions of ground state $A^{Nr}_{0,L/R}(E)$ (continuous) and first excited state $A^N_{1,\alpha}(E)$ reproduced in the STM spectra in b). The STM spectra can also show
signatures of charge neutral excited states shown dotted in a), 
depending on the position of the equilibrium chemical potential (see text). c) Simple schematic depicting the interplay of $A^{Nr}_{0,L/R}(E)$ and  $A^N_{1,L/R}(E)$ resulting in satellite peaks S1 and S2 in conduction spectra of lower right half. d) Conductance spectra reproducing features of both ground and excited state spectral functions $A^{Nr}_{0,L/R}(E)$ and $A^N_{1,\alpha}(E)$. }
\label{fig:spec}
\end{figure}
\item {\it{Excited Spectral Functions}}: In the previous case, the chemical
potential of the left contact is fixed above the transition level
$\epsilon^{Nr}_{00}$ but below $\epsilon^{Nr}_{10}$, thus maintaining the molecule's charge neutrality in its
ground state (i.e., $| N,0 \rangle$), and hence only the Ground State spectral 
signature $A^{N}_{0,j}(E)$ is observed. However, in a general non-equilibrium scenario, 
access to excited states of the neutral
and charged molecule becomes feasible and hence description in terms of spectral
functions corresponding to addition/removal from the $i^{th}$ excited state of
the neutral molecule is required:
\begin{equation}
A^N_{i,\alpha}(E)=\sum_{j}\left[{\tau}^{Nr}_{i j
\alpha}\delta(E-\epsilon^{Nr}_{ij})+{\tau}^{Na}_{ij
\alpha}\delta(E-\epsilon^{Na}_{ij}) \right],
\label{eq:spi}
\end{equation}
where $\alpha$ now corresponds to the two sites that are coupled to the left (L)/
right (R) contacts. For example, let the equilibrium chemical potential be
situated at a position above $\epsilon^{Nr}_{10}=E^{N}_{1}-E^{N-1}_{0}$, shown dotted in Fig.3a.
Given, a positive bias (${\mu}_L>{\mu}_R$) the above transition
is energetically feasible only if the ground state of the cation ($| N-1,0
\rangle$) is accessed, which occurs for a tip voltage corresponding to $\mu_R$ below $\epsilon^{Nr}_{00}$. Once this transition is accessed, spectral function $A^N_{1,L/R}$ involving the first excited state gets involved due to the initial condiction $\mu_L > \epsilon^{Nr}_{10}$, due to which the neutral excited state $| N,1 \rangle$ can be accessed. This results in additional satellite peaks S1 and S2 in Fig. 3d.
A schematic of transitions that consititute the satellite peaks S1 and S2 due to $A^N_{1,\alpha}(E)$ is shown in
Fig.3c. In this figure, we have only shown the removal levels for brevity, and extension of
the argument by including addition levels is trivial. In general, in the STM
regime, one can write a simple expression to evaluate the conductance formula
as a weighted average over various excited state spectral functions:
\begin{equation}
\frac{dI}{dV_R} \approx \frac{e^2}{h}\gamma^{}_R\sum_iP^N_i A_{iR}^{N}(\mu_R).
\end{equation}
\end{itemize}
We have thus shown a simple signature that indicates the access of excitations in the many-body spectrum of
the neutral molecule. In general, one may view near-equilibrium conduction in the CB regime using
single particle energy levels:
\begin{equation}
\epsilon^N_{ij} = E^N_i -E^{N-1}_j
\label{eq:spl}
\end{equation}
and their corresponding spectral
weights.
\subsection{Break Junction limit}
The break junction limit is achieved by setting $\eta=0.5, \gamma=1$, 
implying that both contacts are equally coupled to the molecular
dot and half the applied voltage appears across the molecular levels which in our case transition energies 
$\epsilon^{N}_{ij}$. The many-body configuration of the 
molecule consists of its ground state $|N,0 \rangle$ and the first excited state $|
N,1 \rangle$ separated by a gap similar to the HOMO-LUMO gap $\Delta$, followed by a 
set of closely spaced excitations denoted by $|N,i \rangle, i > 1$. 
The I-V characteristics 
in this limit show certain key signatures which result from how these excitations are accessed.

The onset of conduction is established by the offset between the equilibrium Fermi energy $E_F$ and the 
first accessible transition energy $\epsilon^{Nr}_{00}$. The qualitative shape of the I-Vs depends 
on how the excitations are accessed. Recall that 
$\eta=0.5$ implies that the molecular levels are displaced with respect to the
contact electrochemical potentials by the applied voltage. If the excited states are not accessed 
simultaneously or prior to the threshold
transition $\epsilon^{Nr}_{00}$, as shown in Fig.4b, the I-V has a brief staircase of plateaus 
before a quasi linear rise in current. 
This quasilinear current rise occurs due to a huge number of closely spaced transport 
channels that are triggered only when 
transitions involving an excitation appear within 
the bias window. However, the quasilinear current can also appear prematurely without an intervening plateau, 
if a feasible transition to an excited state appears in the bias window at or before 
the threshold transition.  This situation is shown in Fig.4b, where 
$\epsilon^{Nr}_{10}$ also appears at threshold, resulting in a quasilinear regime immediately
following the onset. 
The two distinct I-Vs have been observed 
experimentally \cite{rdekker,jpark,rweber1,rweber2,pnas} and depend merely on the position of the equilibrium 
electrochemical potential (Fermi energy) with respect to the transition energies.
In the meanwhile, similar SCF based I-V characteristics show adiabatically smeared out currents 
whose onsets get postponed by the changing position of the equilibrium $E_F$, as shown in Fig. 4c. 
The SCF potential from Eq. 5, determines how levels float with respect to their non-equilibrium occupation \cite{rdatta1}. 
It is readily seen by comparing Fig.3a,b with Fig.3c that any self consistent potential cannot change 
the qualitative features of the I-Vs in order to resemble the CB features.
\begin{figure}
\centerline{\epsfig{figure=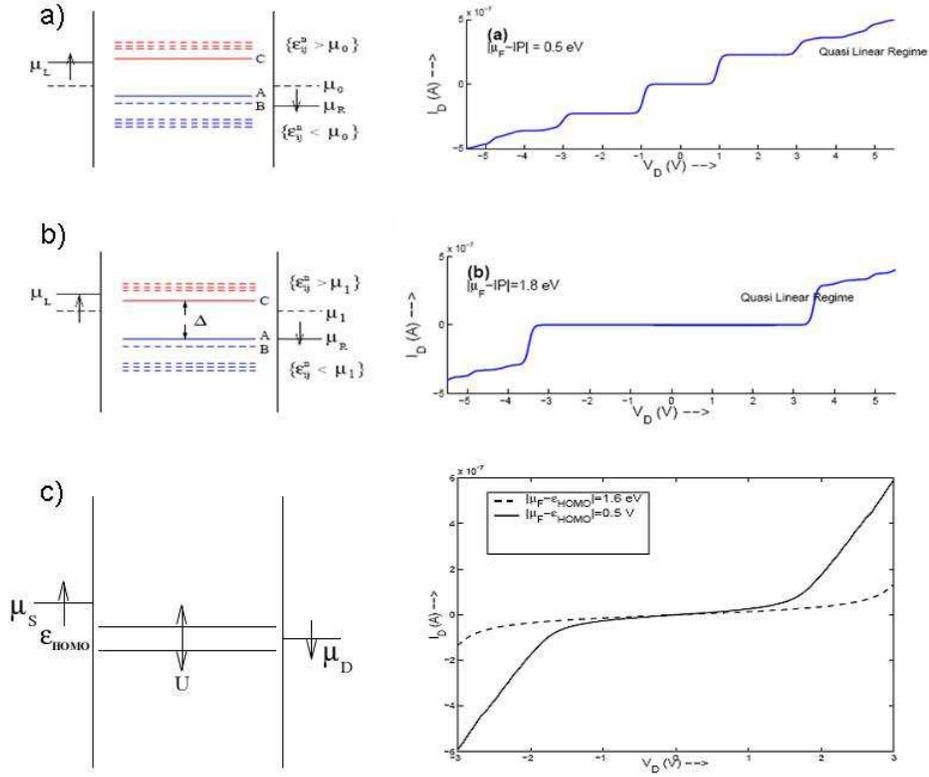,height=4.5in,width=6in}}
%\centerline{\epsfxsize=3.6in\epsfbox{CB_fig4.ps}}
\caption{CB transport under identical contact coupling. a) Schematic of CB conduction resulting in qualitatively different I-V characteristics. For 1) $\mu_F=\mu_0$ one observes I-V with a Coulomb staircase with a plateau followed by a quasi-linear rise. 2) $\mu_F=\mu_1$ one observes the quasi linear I-V upon reaching threshold. (b) This occurs with the intersection of $\mu_L=\epsilon^{Nr}_{10}$ and $\mu_R=\epsilon^{Nr}_{00}$ line in the stability diagram. Stability diagram shown for $N=6$ particle blockade region. c) Distinct I-V's under cross sections $\mu_F=\mu_0$ and $\mu_F=\mu_1$. }
\label{fig:bj}
\end{figure}
\subsection{Connection to experiments: Fitting Data using Coulomb Blockade model}
We consider matching the I-V shapes using our CB model using consistent fitting parameters. 
The experiments conducted on conjugated phenylenes \cite{rweber1,rweber2} at low temperatures ($T=30K$) 
suggest strong Coulomb Blockade effects. It is worth noting that an `orthodox' theory simply involving
junction resistors and capacitors would also manage to capture the zero-bias suppressed conductance, 
the subsequent sharp onset and the linear current rise; however, it would not capture the intervening plateaus, 
fine structures in the I-Vs, and the gateability of the current levels and their asymmetry features that arise
due to discrete transitions in the molecular configuration space \cite{jpark,rscott}. In contrast with 
metallic islands, a molecular dot shows significant size quantization that leads 
to quantum corrections to the junction capacitance, and gets further modified 
at high bias to involve nonlinear corrections to it arising from partial densities of states 
filled separately by the two contacts. 

While our model explains salient features of a lot of Coulomb Blockade experiments \cite{jpark,rscott,rdekker}
it is interesting to note that in some cases the same molecule showed CB behavior at low temperature and SCF behavior at higher
temperatures \cite{rweber1,rweber2}. A possible explanation is that 
at low temperature the molecule could be frozen into a configuration where the
plane of the middle ring is oriented perpendicular to the side rings, 
while room temperature structures sample other configurations and are 
rotated on average. This is supported by the fact that current levels at room temperature are an 
order of magnitude greater, which can be attributed to an increased average degree of 
conjugation along the molecular backbone. 
In contrast at low temperature the rotated central ring has a weaker coupling with the
rest of the backbone, which could reduce its broadening while increasing electron localization and charging,
leading to CB behavior. 
In fact some of the experiments feature bulky middle groups like 
antracence. Steric side groups that are deliberately inserted to facilitate this
rotation of the central rings and enforce CB \cite{rweber2}. While doing exact 
calculations on these molecular structures is beyond the scope of the 
present paper, we consider making simple fits by considering the following facts: 
\begin{itemize}\item Current Levels: Using the fitting parameters
$\gamma_1=\gamma_2 \approx 5-10 meV $ we obtain current levels similar to experimental
data. It is important to note that changing $\gamma$ does not affect
the conductance before the threshold voltage which shows a vanishingly small pre-threshold current.
\item Threshold Voltage: We noticed in the last section with that the 
gap $\Delta$ between ground and first excited states of the neutral molecule 
is important in determining the qualitative shape 
of the I-V. When the equilibrium electrochemical
potential $E_F$ lies above mid-gap between $\epsilon^{Nr}_{10}$ and $\epsilon^{Nr}_{00}$, 
the first excited state becomes voltage-accessible before the ground state of the charged 
species is accessed and populated simultaneously via $\mu_R=\epsilon^{Nr}_{00}$, giving 
rise to the quasi linear I-V immediately following the very first current onset. 
In all the experimental data, we observe a threshold voltage 
between $0.5-0.7$ V thus tuning the gap $\Delta \approx 0.6-0.8$ V.
\end{itemize}
\begin{figure}
\centerline{\hskip 1cm\epsfig{figure=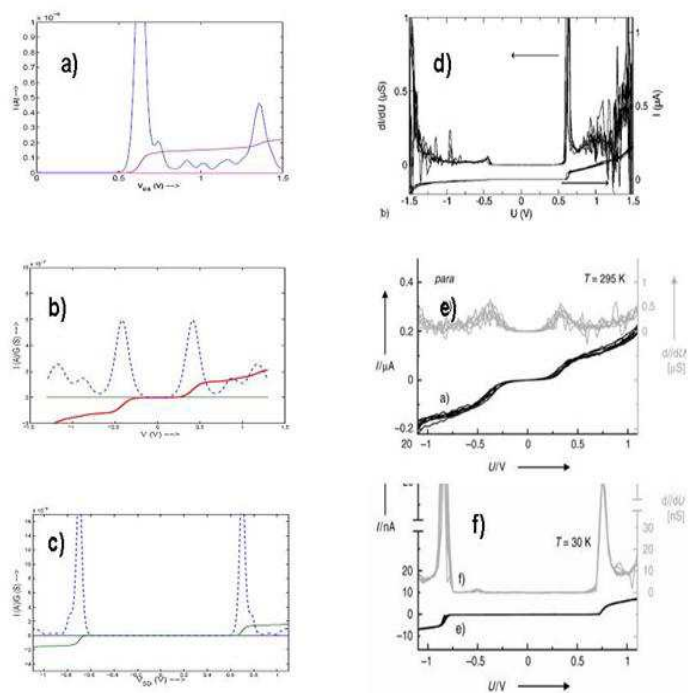,height=5in,width=6in}}
%\centerline{\epsfxsize=4in\epsfbox{CB_fig1a.ps}}
\caption{Experimental fits for data \cite{rweber1,rweber2}. a) $T=30K$, $\gamma \approx 5meV$. b) $T=295K$, $\gamma \approx 5meV$ c)$T=30K$, $\gamma \approx 0.25 meV$. }
\label{fig:wfit}
\end{figure}
Fig.~\ref{fig:wfit}(a) (b) and (c) are fits obtained for experiments \cite{rweber1,rweber2}.
In case of molecular asymmetries \cite{rweber1}, only positive bias is considered \cite{rbhasko}, 
the I-V asymmetries themselves being attributed to polarization 
effects \cite{rasymm}. In obtaining an experimental fit for \cite{rweber2}, in
Fig.~\ref{fig:wfit}(b), we used $T=295$ K consistent 
with experiment. Notice that the first peak has broadened significantly.
The higher temperature Coulomb Blockade could possibly be attibuted to the fact that molecule involved has  an anthracene based middle ring that is much bulkier, thus leading to a 
higher temperature frozen configuration stabilized by steric 
interactions.

The molecular system we consider is a simple prototypical molecule 
(benzene in our case) with calculations based on
simple parameters that are associated with
this minimal system. Performing calculations on a 
real molecule-electrode system will be needed to
yield a quantitative fit in terms of threshold voltage, current levels and
positions of peaks. However, the conduction mechanism remains the same. The exponentially
larger configuration space of even a minimal Coulomb Blockaded molecule makes a first-principles calculation of its transport
properties inordinately challenging compared to SCF treatments in the literature. However, the SCF calculations
do not capture the non-equilibrium transition rates between the many-body states, which as we argued earlier
carry crucial correlation signatures that are experimentally observable for ultrashort molecules. 
Such a ``real'' calculation involving the quantum chemistry of larger molecules and 
contact bondings within this nonequilibrium full CI treatment 
is still at a very early stage \cite{rdelaney2}. Furthermore, it needs to be supplemented
with the broadening of the many-particle states that could affect the interference between nearby levels, 
an issue that has received relatively little attention \cite{rjauho,rbraig,mwl2,rgurevich,aw} 
and requires further work.
\section{conclusion}
In this paper, we have developed a Coulomb Blockade approach for molecular conduction through 
short molecules using Benzene as a prototype. We have shown how equilibrium and non-equilibrium 
signatures are very different from the traditional NEGF-SCF viewpoint, and that the CB approach 
is appropriate in the weak coupling limit. Many I-V features distinct to the CB regime are often seen 
in experiments. These features that are easily obtained using a full Configuration Interaction 
master equation approach are potentially very hard to obtain within any 
effective one-electron potential, even for a minimal model. A particular challenge
therefore lies in bridging the SCF and CB regimes while paying close attention to coherent level
broadening and associated interferences. The emergence of many recent experiments on molecular dots,
exploring the interplay between charging, quantization and level-broadening, should prove invaluable
in further theoretical developments in this regard.

%\begin{biography}[{\includegraphics[width=1in,height=1.25in,clip,keepaspectratio]{mshell}}]{Michael Shell}
% where an .eps filename suffix will be assumed under latex, and a .pdf suffix
% will be assumed for pdflatex; or if you just want to reserve a space for
% a photo:

%\begin{biography}{Michael Shell}
%\end{biography}

%\begin{biographynophoto}{John Doe}
%\end{biographynophoto}

% insert where needed to balance the two columns on the last page
%\newpage

%\begin{biographynophoto}{Jane Doe}
%\end{biographynophoto}


%\vfill


\end{document}